\documentclass[fleqn,usenatbib]{mnras}

\usepackage{newtxtext,newtxmath}
\usepackage[T1]{fontenc}
\usepackage{ae,aecompl}

\usepackage{graphicx}	% Including figure files
\usepackage[dvipsnames]{xcolor}	% Including figure files
\usepackage{amsmath}	% Advanced maths commands
\usepackage{amssymb}	% Extra maths symbols

\usepackage{savesym}
\savesymbol{tablenum}
\usepackage{siunitx}
\restoresymbol{SIX}{tablenum}

\newcommand{\uf}{$u_{\rm f}$}
\newcommand{\mo}{M$_\odot$}
\newcommand{\me}{M$_\oplus$}
\newcommand{\ms}{m s$^{-1}$}
\newcommand{\es}{erg s$^{-1}$}
\newcommand{\lx}{$L_{\rm X}$}
\newcommand{\yr}{yr$^{-1}$}
\newcommand{\md}{$\dot{M}_{\rm wind}$}

\newcommand{\mpl}{$M_{\rm pl}$}
\newcommand{\sd}{$\dot{\Sigma}_{\rm wind}$}
\newcommand{\gsig}{$\Sigma_{\rm g}$}
\newcommand{\dsig}{$\Sigma_{\rm d}$}

\title[X-ray photoevaporation and the streaming instability]{X-ray photoevaporation's limited success in the formation of planetesimals by the streaming instability}

\author[Ercolano, Jennings, Rosotti, Birnstiel]{
Barbara Ercolano$^{1,2}$, Jeff Jennings$^{1}$, Giovanni Rosotti$^{3}$, Tilman Birnstiel$^{1}$
\\
$^{1}$University Observatory, Faculty of Physics, Ludwig-Maximilians-Universit\"at M\"unchen, Scheinerstr. 1, 81679 Munich, Germany\\
$^{2}$Excellence Cluster Origin and Structure of the Universe, Boltzmannstr.2, 85748 Garching bei M{\"u}nchen, Germany\\
$^{3}$Institute of Astronomy, University of Cambridge, Madingley Road, Cambridge, UK
\\
}

\date{Accepted XXX. Received YYY; in original form ZZZ}

\pubyear{2017}

\begin{document}
\label{firstpage}
\pagerange{\pageref{firstpage}--\pageref{lastpage}}
\maketitle

\begin{abstract}
The streaming instability is often invoked as solution to the fragmentation and drift barriers in planetesimal formation, catalyzing the aggregation of dust on kyr timescales to grow km-sized cores. However there remains a lack of consensus on the physical mechanism(s) responsible for initiating it. One potential avenue is disc photoevaporation, wherein the preferential removal of relatively dust-free gas increases the disc metallicity. Late in the disc lifetime, photoevaporation dominates viscous accretion, creating a gradient in the depleted gas surface density near the location of the gap. This induces a local pressure maximum that collects drifting dust particles, which may then become susceptible to the streaming instability. Using a one-dimensional viscous evolution model of a disc subject to internal X-ray photoevaporation, we explore the efficacy of this process to build planetestimals. Over a range of parameters we find that the amount of dust mass converted into planetesimals is often < 1 \me\ and at most a few \me\ spread across tens of AU. We conclude that photoevaporation may at best be relevant for the formation of debris discs, rather than a common mechanism for the formation of planetary cores. Our results are in contrast to a recent, similar investigation that considered an FUV-driven photoevaporation model and reported the formation of tens of \me\ at large (> 100 AU) disc radii. The discrepancies are primarily a consequence of the different photoevaporation profiles assumed. Until observations more tightly constrain photoevaporation models, the relevance of this process to the formation of planets remains uncertain.
\end{abstract}


%%%%%%%%%%%%%%%%%%%%%%%%%%%%%%%%%%%%%%%%%%%%%%%%%%

%%%%%%%%%%%%%%%%% BODY OF PAPER %%%%%%%%%%%%%%%%%%
\section{Introduction}

The growth of dust grains to build planetesimals, km-size bodies that will subsequently form planetary cores, is known to be theoretically problematic.  On the many orders of magnitude journey from micron to km sizes, dust particles encounter a number of growth barriers that suggest the process is largely inefficient. The fragmentation barrier and the radial drift barrier are prominent examples, affecting particles that approach Stokes numbers of $\approx 0.1$ \citep{2008A&A...480..859B,2010A&A...513A..79B,2012A&A...539A.148B}. The ubiquity of planets implies these barriers are commonly overcome, either via collisional process \citep[such as fractal growth, see][]{ 2012ApJ...752..106O,2013A&A...557L...4K} or possibly via mechanisms that trap the particles in gas pressure maxima within the protoplanetary disc \citep{2004A&A...425L...9P,2006MNRAS.373.1619R,2012A&A...545A..81P,2012ApJ...755....6Z}. Such \lq{}traps\rq{} could be induced by, e.g., vortices \citep[e.g.,][]{1995A&A...295L...1B,1997Icar..128..213K,2013A&A...550L...8B,2013ApJ...775...17L,2016MNRAS.458.3927B}, snowlines \citep[e.g.,][]{2007ApJ...664L..55K}, or dynamical structures such as spiral arms \citep{Rice2004,Rice2006,Gibbons2012,Gibbons2014,Dipierro2015,Booth2016}.

In this paper we explore the possibility that gas removal in these discs due to photoevaporation may provide suitable conditions to trigger collective mechanisms for planetesimal formation such as the streaming instability \citep[e.g.,][]{2010AREPS..38..493C}. Photoevaporative winds could help in two ways: (i) they preferentially remove gas from the disc (only the smallest ($\sim \mu{\rm m}$-size) dust particles are entrained in the wind, \citealt{2011MNRAS.411.1104O}), hence enhancing the dust-to-gas ratio in the disc; (ii) the mass loss is concentrated in a limited range of disc radii, creating surface density gradients and thus pressure maxima in the disc that may act as dust traps.

A review of previous studies on the role of photoevaporation in planet formation can be found in \citet{2014prpl.conf..475A} and \citet{2017RSOS....470114E}; only the main results are summarised here. \citet{2005ApJ...623L.149T} explored the case of external EUV/FUV-driven photoevaporation of protoplanetary discs and found that this yielded a significant enhancement of the dust-to-gas ratio between $5 - 50$ AU, with values high enough to reach the threshold for gravitational instability as given by  \citet{2002ApJ...580..494Y}. \citet{2007MNRAS.375..500A} used a 1-D model of EUV-driven photoevaporation to show that a radial pressure gradient can lead to the formation of a ring in which the dust-to-gas ratio is enhanced, as gas disc dispersal proceeds from the innermost radii outward. However as noted in \citet{2014prpl.conf..475A}, the relevance of this process to the formation of giant planets is likely limited because EUV-driven photoevaporation, which has mass loss rates of at most \SI{e-10}{M_\odot.yr^{-1}}, only affects the viscous evolution late in a disc's lifetime, after most of the gas has already been accreted onto the star. Compounding this effect, the amount of solids remaining in the later stages of disc evolution (when EUV-driven photoevaporation becomes relevant), particularly at large disc radii, may be significantly depleted by radial drift \citep{2005ApJ...627..286T, 2012MNRAS.423..389H} unless an efficient particle trapping mechanism operates \citep[e.g.,][]{2009A&A...503L...5B,2012A&A...538A.114P}.

\citet{2017ApJ...839...16C} recently presented a study of planetesimal formation by the streaming instability (SI) in the context of FUV-driven photoevaporation using the models of \citet{2015ApJ...804...29G}. They are able to produce massive ($60-130$ {\me}) planetesimal belts beyond 100 AU and up to 20 {\me} between $3-100$ AU, as well as 8 {\me} interior to 3 AU by additionally invoking a dust trap. These results are in stark contrast to previous studies and if confirmed would imply that photoevaporation plays an appreciable role in planet formation. The FUV-driven photoevaporation models of \citet{2015ApJ...804...29G} produce a significantly more vigorous wind than models driven by EUV radiation, but the absolute wind rates and profile depend on a number of assumptions that influence the heating efficiency of the FUV photons in the disc atmosphere, e.g., the abundance of polycyclic aromatic hydrocarbons (PAHs), which are rarely  detected in T-Tauri discs \citep{2006A&A...459..545G}. The largest source of uncertainty in these models is that they are not based on hydrodynamic calculations; the surface mass loss rates beyond an analytically determined gravitational radius in a hydrostatic disc structure are estimated by taking the maximum of $\Sigma_{\rm g} c_{\rm s}$ along the vertical direction \citep{2009ApJ...690.1539G}, where {\gsig} is the local gas disc surface density and $c_{\rm s}$ the isothermal sound speed. The uncertainties introduced by this approach have yet to be fully explored.

\begin{figure}
\includegraphics[width=0.47\textwidth]{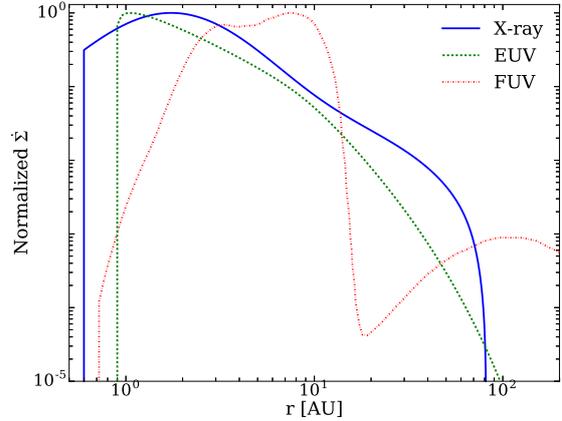}
\caption{Normalized radial photoevaporative mass loss profiles for common X-ray-, EUV- and FUV-dominated models, adapted from \citet{2014prpl.conf..475A} Figure 3.}
\label{fig:profiles}
\end{figure}

For X-ray-driven photoevaporative winds, radiation hydrodynamic solutions have been calculated for the case of a solar-type star \citep{2010MNRAS.401.1415O,2011MNRAS.412...13O,2012MNRAS.422.1880O}, and these detailed calculations have shown that X-rays are able to drive much more vigorous winds than EUV radiation. Typical solar-type stars with X-ray luminosities of $\sim \SI{e30}{erg.s^{-1}}$ drive mass loss rates of $\sim \SI{e-8}{M_\odot.yr^{-1}}$ \citep{2010MNRAS.401.1415O}, the same order of magnitude as estimates for FUV-driven mass loss rates. X-ray photoevaporation rates and profiles have been obtained by means of radiation-hydrodynamic simulations and have been shown to be insensitive to  heating processes (e.g., FUV) occurring in the layer below the X-ray penetration front \citep{2012MNRAS.422.1880O}. While the total mass loss rates obtained for solar-type stars with a median X-ray luminosity of $\sim 10^{30}$ erg s$^{-1}$ are comparable to some FUV-driven models from \citet{2015ApJ...804...29G}, the wind profiles obtained are significantly different, which has important consequences for planetesimal formation by the streaming instability as we show in this paper. A comparison of typical wind profiles for the X-ray-, EUV- and FUV-driven cases is shown in the reviews by \citet{2014prpl.conf..475A} and \citet{2011ARA&A..49..195A}, and reproduced in Figure~\ref{fig:profiles} of this paper. 

The relevance of photoevaporation in forming planetesimals by this process is explored in this work with a 1-D viscous evolution model of a protoplanetary disc subject to X-ray and EUV photoevaporation from its host star, including a prescription to self-consistently treat the evolution of the dust disc. The methods employed are described in Section 2, while the results are presented in Section 3 and discussed in Section 4. Section 5 contains a brief summary of our work and conclusions.

\section{Strategy and Methods}
\label{sec:methods}

To zeroth order, a strong concentration of solids can be obtained by the SI if a \lq{}metallicity\rq{} threshold of $Z \geq Z_{\rm crit} \approx 0.015 - 0.020$ is met, where $Z$ is the local surface density ratio of large pebbles to gas. \citet{2014A&A...572A..78D} show that $Z_\mathrm{crit}$ strongly depends on particle size, or more precisely on the dimensionless stopping time (Stokes number, St). For metallicities $\gtrsim 10^{-2}$, \cite{2014A&A...572A..78D} show that Stokes numbers of order $10^{-2}$ are sufficient to trigger the instability, and \cite{2015A&A...579A..43C} extend the condition down to Stokes numbers of order $10^{-3}$ for similar $Z$. The exact critical values for $Z$ and St are subject to further complications; simulations show that for larger pressure gradients (higher drift speeds), the metallicity threshold for clumping increases \citep{2010ApJ...722L.220B}, and this effect is not taken into account in the estimates above. Moreover, in pressure maxima (where the pressure gradient vanishes) the threshold may instead decrease, but it is currently unclear how the clumping would operate in these cases, as some drift is needed to drive the SI. Finally, most simulations to-date assume a quiet midplane, apart from stirring caused by the SI and associated particle-gas interactions. If additional turbulence is present, larger particles are probably necessary for the SI to operate.

While these processes may appreciably influence the $Z$ and St thresholds needed to trigger the SI, we neglect them here because of the current high uncertainties in the magnitude of their effects. In this work we adopt the criteria identified by \cite{2015A&A...579A..43C} to assess the relevance of photoevaporation in the formation of planetesimals, using a 1-D model of a viscously evolving disc subject to X-ray and EUV-driven photoevaporation \citep{2015MNRAS.450.3008E} from a 0.7 {\mo} central star. We use the two-population model of \citet{2012A&A...539A.148B} as a prescription for the evolution of dust particles, investigating the distribution of $Z$ and St in the disc as it evolves and ultimately disperses. We follow \citet{2017ApJ...839...16C} to determine the amount of material forming planetesimals at each radius in the disc as a function of time under the criteria described in Section~\ref{sec:sicrit} for triggering the SI. At the times and locations where the SI occurs, we immediately remove 90\% of the local dust. To explore the sensitivity of our results, we vary model input values for the dust particle fragmentation velocity {\uf}, the viscosity parameter $\alpha$ and initial disc scaling radius $R_1$ \citep{1974MNRAS.168..603L}, the starting time for the evolution of the dust $t_{\rm delay}$, and the X-ray luminosity {\lx} of the central star. Table~\ref{tab:models} provides a summary of the input parameters for each model.

\begin{table*}
\noindent
\begin{tabular}{lcccccc}
\hline
Model           & {\lx} [$10^{30}$ {\es}]                      & {\md} [$10^{-9}$ {\mo} yr$^{-1}$] & {\uf} [m s$^{-1}$]  & $t_{\rm delay}$ [Myr] & $R_1$ [AU] & $\alpha$ \\
Fiducial & 1.0 & 6.4 & 10 & 0 & 18 & $7 \times 10^{-4}$\\
Uf5 & 1.0 & 6.4 & 5 & 0 & 18 & $7 \times 10^{-4}$\\
Uf35 & 1.0 & 6.4 & 35 & 0 & 18 & $7 \times 10^{-4}$\\
Lx0.1 & 0.1 & 0.5 & 10 & 0 & 18 & $7 \times 10^{-4}$\\
Lx0.5 & 0.5 & 2.9 & 10 & 0 & 18 & $7 \times 10^{-4}$\\
Lx5 & 5.0 & 40.2 & 10 & 0 & 18 & $7 \times 10^{-4}$\\
Lx10 & 11.0 & 88.5 & 10 & 0 & 18 & $7 \times 10^{-4}$\\
Td1 & 1.0 & 6.4 & 10 & 1 & 18 & $7 \times 10^{-4}$ \\
Td2 & 1.0 & 6.4 & 10 & 2 & 18 & $7 \times 10^{-4}$\\
Ri200 & 1.0 & 6.4 & 10 & 0 & 200 & 0.01\\
\hline
\end{tabular}
\caption{Summary of model parameters for a disc subject to X-ray photoevaporation from its 0.7 {\mo} central star. Model names give the parameter value that differs from the fiducial case: dust fragmentation velocity in {\ms} (Uf: 5 or 35, as opposed to the fiducial 10), stellar X-ray luminosity in $10^{30} $ {\es} (Lx: 0.1, 0.5, 5.0, or 10.0), delayed start of the dust disc evolution in Myr (Td: 1 or 2), or initial disc radius (Ri: 200 AU instead of the fiducial 18 AU) and viscosity parameter (0.01 in place of the fiducial $7 \times 10^{-4}$).}
\label{tab:models}
\end{table*}

In the next sub-section the criteria adopted for the triggering of the streaming instability are briefly summarised. Our viscous evolution code and our implementation of a two-population prescription to model dust evolution are described in the subsequent two subsections. 

\subsection{Streaming instability criteria}
\label{sec:sicrit}
To trigger the SI we use the criteria introduced by \citet{2017ApJ...839...16C}: Stokes numbers and metallicity must be larger than critical values, St$_{\rm crit}$ = 0.003 and $Z_{\rm crit}$ = max($Z_1$, $Z_2$), with
\begin{equation}
\label{eq:zcrit}
Z_1 = \sqrt{\frac{\alpha_{\rm t}}{{\rm St}+\alpha_{\rm t}}}; Z_2 = 10^{-1.86+0.3(0.98+{\rm log_{10}St})^2}.
\end{equation}
We follow \citet{2017ApJ...839...16C} and assume stratified turbulence, whereby $\alpha_{\rm t}$ represents the value at the midplane and is taken a factor of 100 lower than the global $\alpha$ used for viscosity calculations.  For the X-ray photoevaporation model implemented in our work $\alpha = 7 \times 10^{-4}$, and thus  $\alpha_{\rm t} = 7 \times 10^{-6}$. This is a very low value, and thus the results presented here represent an optimisic case for planetesimal formation. We also run experiments using the refined, less stringent criteria for triggering the SI in Figure 9 of \citet{2016arXiv161107014Y}, however the resulting increase in planetesimal production is minimal, and so we limit subsequent discussion to the results obtained using the \citet{2017ApJ...839...16C} criteria.

\subsection{Evolution of the gas disc}
The system evolves according to
\begin{equation}
\frac{\partial \Sigma_{\rm g}}{\partial t} = \frac{1}{r} \frac{\partial}{\partial r} \bigg[ 3r^{1/2} \frac{\partial}{\partial r}\big(\nu \Sigma_{\rm g} r^{1/2}\big)\bigg] - \dot{\Sigma}_{\rm wind}(r,t),
\label{eq:evo}
\end{equation}
where the first term on the right-hand side describes the viscous evolution of the disc \citep{1974MNRAS.168..603L} and the second the mass loss due to photoevaporation \citep[e.g.,][]{2001MNRAS.328..485C}. $\Sigma_{\rm g}$ is the gas disc surface density, $r$ the radial distance from the star in the disc midplane, $\nu$ the kinematic viscosity of the disc, $M_*$ the stellar mass, and {\sd} the radial photoevaporation profile.

To solve Equation~(\ref{eq:evo}), we use the 1D viscous evolution code $\textsc{SPOCK}$ detailed in \cite{2015MNRAS.450.3008E}. We discretize Equation~(\ref{eq:evo}) on a grid of 1000 points equispaced in $r^{1/2}$ between $0.04 - 10^4$ AU. We prescribe $\nu=\alpha c_{\rm s} H$ \citep{ShakuraSunyaev}, where $c_{\rm s}$ is the sound speed and $H$ the disc scale height. We assume a disc temperature structure $T \propto r^{-1/2}$, with $T \approx 2100$ K and 4 K at the inner and outer boundaries, respectively. Although throughout the text we refer only to the value of $\alpha$, note that the physical quantity in our equations for the gas is only the kinematic viscosity $\nu$; the values of $\alpha$ that we use are therefore degenerate with the values of the temperature adopted. To integrate the viscous term in Equation~(\ref{eq:evo}), we perform a change of variables to recast the equation into a diffusion equation \citep{1986MNRAS.221..169P}. The photoevaporation term is integrated by decreasing $\Sigma$ at every timestep by the amount $\dot{\Sigma} \Delta t$, where $\Delta t$ is the length of the timestep. We refer to the Appendix of \citet{2012MNRAS.422.1880O} for the value of " $\dot{\Sigma}$ in the X-ray photoevaporation case. To prevent numerical problems, we use a floor surface density of $10^{-8}$ g cm$^{-2}$.

To treat photoevaporation, we use the X-ray dominated model in \citet{2010MNRAS.401.1415O} that includes a secondary EUV component and is derived using a hydrodynamic solution for the wind. For a 0.7 {\mo} star, the integrated gas mass loss rate across the disc is {\md} $= \int 2 \pi r^2 {\dot \Sigma}_{\rm wind} (r)\ \mathrm{d}r \approx 7 \times 10^{-9}$ {\mo} {\yr}. The X-ray model has two epochs delineated by the clearing of a hole in the disc, at which point the inner edge of the outer disc is exposed directly to stellar irradiation. Assuming an X-ray penetration depth of $10^{22}$ cm$^{-2}$ \citep{Ercolano2009}, we switch to the second epoch once the hydrogen column density is below this value out to the location of gap opening. The simulation ends when the photoevaporative hole reaches 100 AU.

\subsection{Evolution of the dust disc}
In treating the dust disc we take into account the radial drift induced by the pressure gradient of the gas, turbulent diffusion and the effect of collisions leading to grain growth and fragmentation. We evolve the system as
\begin{equation}
\frac{\partial \Sigma_\mathrm{dust}}{\partial t} = \frac{1}{r} \frac{\partial}{\partial r} \left[R \Sigma_\mathrm{dust} v_\mathrm{drift} - D R \Sigma_\mathrm{g} \frac{\partial}{\partial r} \left( \frac{\Sigma_\mathrm{dust}}{\Sigma_\mathrm{g}}\right) \right],
\label{eqn:devo}
\end{equation}
where the first term describes radial drift, with $v_\mathrm{drift}$ the radial drift velocity of the dust; and the second term turbulent diffusion \citep{1988MNRAS.235..365C,2010A&A...513A..79B}, where $D$ is the dust diffusion coefficient, which we assume equal to the kinematic viscosity $\nu$ of the gas. To solve Equation~\ref{eqn:devo} we use the same grid and change of variables as for the gas disc. We treat the diffusive term as an advection term since the diffusive velocity is typically smaller than the radial drift velocity, and we use a standard upwind method to solve for advection; to reconstruct the values at the cell interface we use the \citet{1977JCoPh..23..276V} method. To compute the dust drift velocity, we follow \citet{TakeuchiLin2001},
\begin{equation}
v_\mathrm{drift} = \frac{{\rm St}^{-1} v_\mathrm{g} - \eta v_\mathrm{k}}{{\rm St+St^{-1}}},
\end{equation}
where $v_\mathrm{g}$ is the radial velocity of the gas induced by accretion and $v_\mathrm{k} = (G M_\ast/r)^{1/2}$ the Keplerian velocity. $\eta$ measures the importance of the pressure gradient with respect to gravity and is given by
\begin{equation}
\eta = - \left(\frac{H}{R}\right)^2 \left(\frac{{\rm d} \log \Sigma_{\mathrm g}}{{\rm d} \log R} + (q-3) \right),
\end{equation}
where $H/R$ is the disc aspect ratio and $q = 5/4$ assuming a standard flaring disc. ${\rm St}=\pi a \rho_{\rm d}/2 \Sigma_{\rm g}$ is the Stokes number of a dust grain of radius $a$ and bulk density $\rho_{\rm d}$, which we assume to be 1 g cm$^{-3}$.

To model dust growth and fragmentation, we follow the simplified prescriptions of \citet{2012A&A...539A.148B} that allow us to consider only one dust species at each radial location, greatly simplifying the computational problem. We start from grains with size of 0.1 $\mu$m that grow on a timescale $t_\mathrm{growth} = (\epsilon \Omega)^{-1}$, where $\epsilon$ is the dust-to-gas ratio $Z$. Further growth is limited by either radial drift or fragmentation. Drift prevents grains from growing beyond a maximum size of
\begin{equation}
a_\mathrm{drift}=0.55 \frac{2 \Sigma_\mathrm{d}}{\pi \rho_{\rm d}} \frac{v^2_{\rm k}}{c_{\rm s}^2} \left|  \frac{\mathrm{d} \log P}{\mathrm{d} \log R} \right|^{-1},
\end{equation}
where $P$ is the gas pressure. Fragmentation sets a maximum grain size of
\begin{equation}
a_\mathrm{frag}=0.37 \frac{2}{3 \pi} \frac{\Sigma_{\rm g}}{\rho_{\rm d} \alpha} \frac{u^2_{\rm f}}{c_{\rm s}^2},
\end{equation}
where $u_f$ is the fragmentation velocity of the dust. We assume that a fraction of the dust population is comprised of grains of this size, while there is always a population of very small grains replenished by fragmentation. If the maximum grain size is limited by radial drift, we assume that 97\% of the grains are large; if in the fragmentation dominated regime, this is instead 75\%. We use these fractions to compute a weighted average of the radial drift velocity to advect the dust (see \citealp{2012A&A...539A.148B} or \citealp{2016SSRv..205...41B} for a review).

\section{Results}

\begin{figure*}
\includegraphics[width=\textwidth]{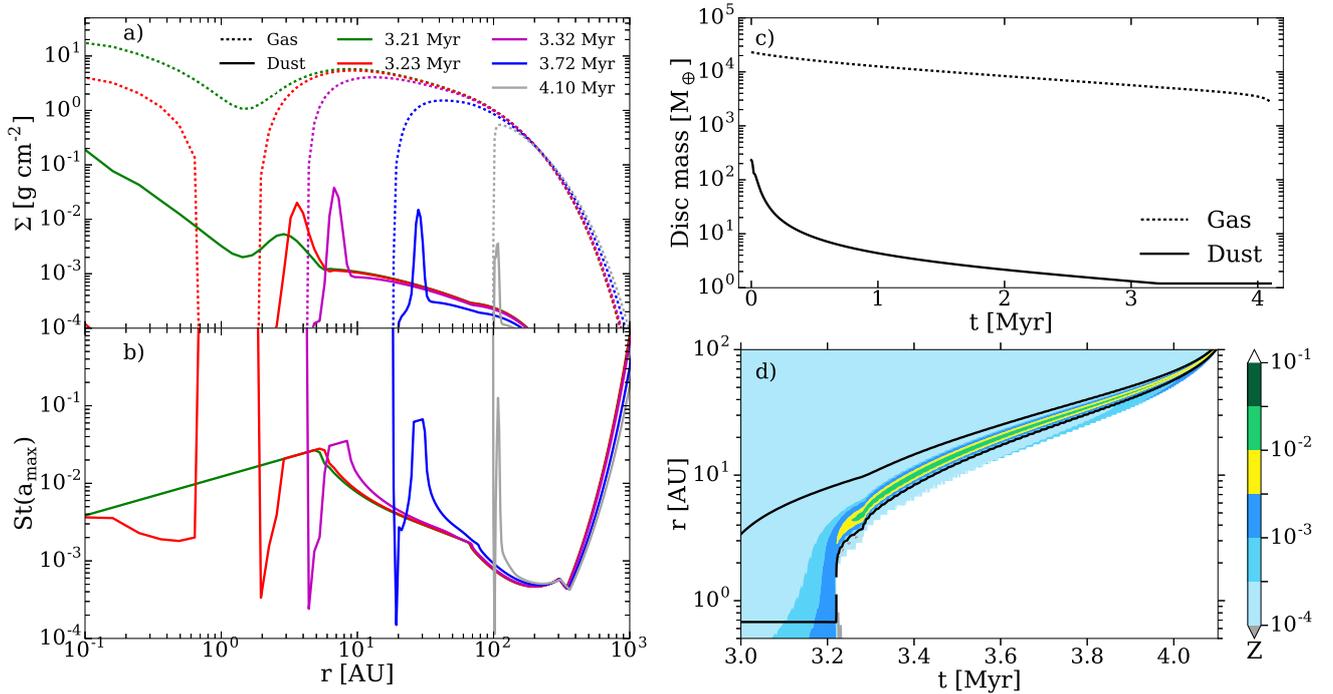}
\caption{Evolution of the fiducial X-ray model Uf10, with parameters given in Table~\ref{tab:models}. a) Disc surface density 10 kyr prior to a photoevaporative gap opening at 3.22 Myr (gas, {\gsig}, green dashed; dust, {\dsig}, green solid), 10 kyr after gap opening (red), 100 kyr after gap opening (purple), 500 kyr after gap opening (blue), and at the time when the inner cavity has reached a radius of 100 AU ($t = 4.10$ Myr; gray). Note the continued retention of the pressure bump along the leading edge of the photoevaporative gap/hole. 
b) Stokes number for dust particles of the largest grain size at the epochs in (a).
c) Gas and dust disc masses over time. A significant amount of dust drifts inward from large radii at early times, depleting the dust surface density for later conversion into planetesimals (see Sec.~\ref{sec:discussion}).
d) Evolution of the dust-to-gas ratio $Z$ for the final $\approx 25\%$ of the gas disc lifetime and out to 100 AU in the disc. $Z$ values are given by the colorbar at right. The white region corresponds to the interior of the gap and subsequent hole in the gas disc induced by photoevaporation. The pressure bump in (a) is seen here as the narrow region in green and yellow. The area interior to the overlaid black contour has Stokes values 0.01 < St < 0.10. The contour's hard lower cut signifies the boundary between the drift-dominated (above) and fragmentation-dominated (below) regimes.
}
\label{fig:fiducial}
\end{figure*}

\begin{figure*}
\includegraphics[width=\textwidth]{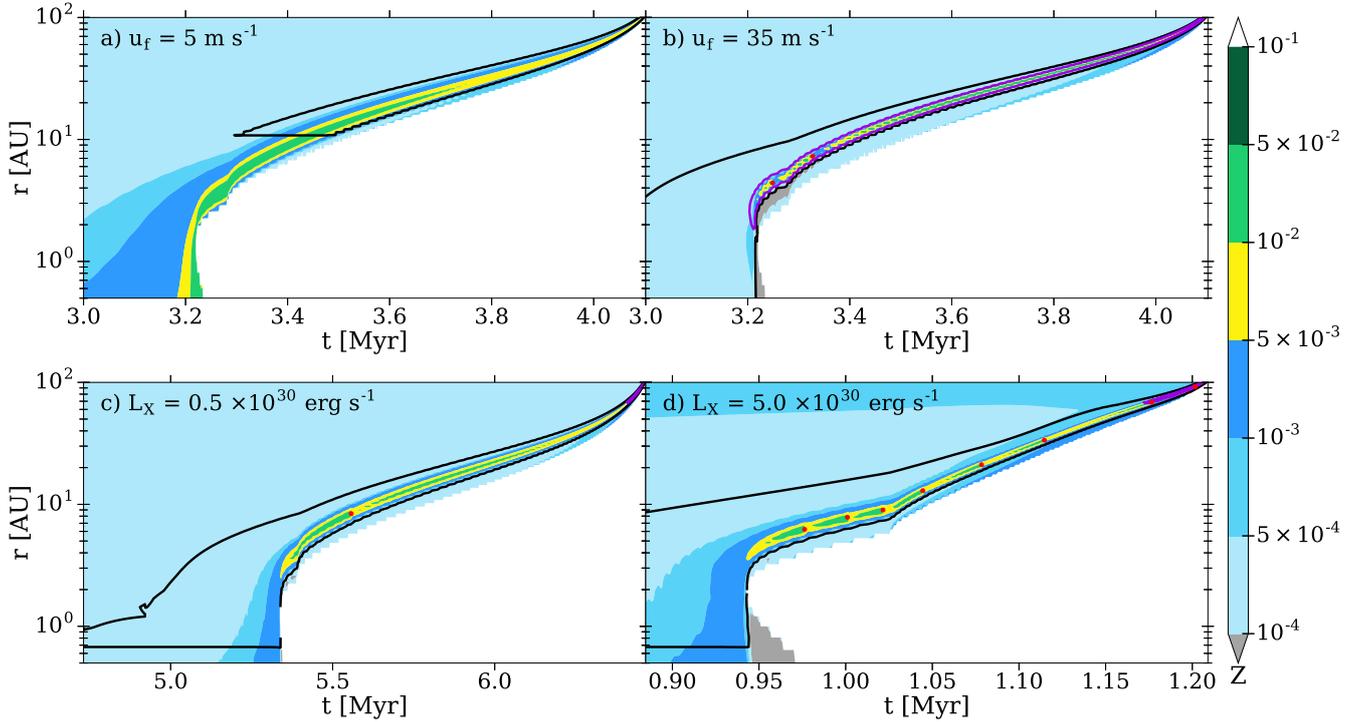}
\caption{Evolution of the dust-to-gas ratio Z for the final $\approx 25\%$ of the gas disc lifetime and out to 100 AU in the disc. The four panels show models in which only one parameter is changed relative to the fiducial case shown in Fig.~\ref{fig:fiducial}(d). Z values are given by the colourbar at right. The white region corresponds to the interior of the gap and subsequent hole in the gas disc induced by photoevaporation. a) Model Uf5, with a fragmentation velocity {\uf} = 5 {\ms}. The result is a higher dust-to-gas ratio but lower Stokes number in the pressure bump.
b) Model Uf35 for {\uf} = 35 {\ms}. Interior to the purple contour, $0.1 < {\rm St} \leq 1.0$. Red points denote times and locations at which the streaming instability is triggered according to the criteria in Section~\ref{sec:sicrit}.
c) Model Lx0.5 for a factor of 5 smaller {\lx} than the fiducial model. The reduced stellar X-ray luminosity takes longer to open a photoevaporative gap, yet also requires longer to disperse the disc, allowing slightly higher dust concentrations in the pressure bump.
d) Model Lx5 for a factor of 5 larger {\lx} than the fiducial model. Because the photoevaporative gap is opened quickly, the dust disc mass is still relatively high late in the disc lifetime, resulting in multiple streaming instability events.}
\label{fig:surface}
\end{figure*}

\begin{figure*}
\includegraphics[width=\textwidth]{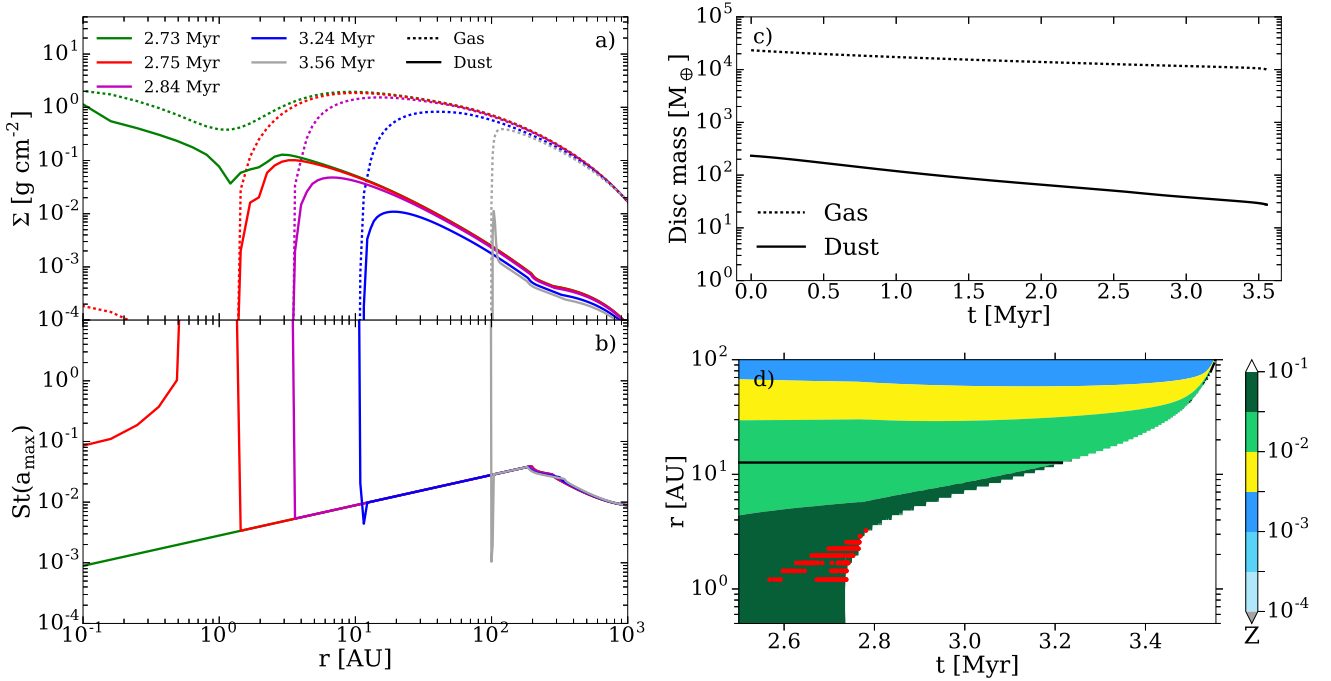}
\caption{Evolution of the Ri200 model (see Table~\ref{tab:models}), in which the viscosity coefficient $\alpha$ at the disc midplane is increased to 0.01 (cf. $\alpha \approx 7 \times 10^{-4}$ in the fiducial model) and the initial disc radius increased to 200 AU (cf. 18 AU in the fiducial model) to maintain the fiducial model's viscous time. Subplots (a) - (c) are scaled in the vertical as in Figure~\ref{fig:fiducial}. a) Disc surface density 10 kyr prior to a photoevaporative gap opening at 2.74 Myr (gas, {\gsig}, green dashed; dust, {\dsig}, green solid), 10 kyr after gap opening (red), 100 kyr after gap opening (purple), 500 kyr after gap opening (blue), and at disc dispersal ($t = 3.56$ Myr; gray).
b) Stokes number for dust particles of the largest grain size at the epochs in (a).
c) Gas and dust disc masses over time. Contrasted with the fiducial model in Figure~\ref{fig:fiducial}(c), the larger initial disc radius used here reduces the dust drift timescale in the outer disc significantly, resulting in a larger dust surface density {\dsig} and thus dust-to-gas ratio $Z$ at late times to trigger the streaming instability (see Sec.~\ref{sec:discussion}).
d) Evolution of $Z$ showing the period over which photoevaporation becomes important in the disc evolution. $Z$ values are given by the colourbar at the right. The white region corresponds to the interior of the gap and subsequent hole in the gas disc induced by photoevaporation. The region below the overlaid black contour has Stokes values 0.01 < St < 0.10. The contour's hard cut signifies the boundary between the drift-dominated (above) and fragmentation-dominated (below) regimes. Red points denote times and locations at which the streaming instability is triggered according to the criteria in Section~\ref{sec:sicrit}. Note these events occur not in a prominent pressure bump as in the fiducial model, but near the depression in the gas surface density that is excavated as photoevaporation works to open a gap (see Section~\ref{sec:ri200}).
}
\label{fig:hialpha}
\end{figure*}

\begin{figure*}
\includegraphics[width=.65\textwidth]{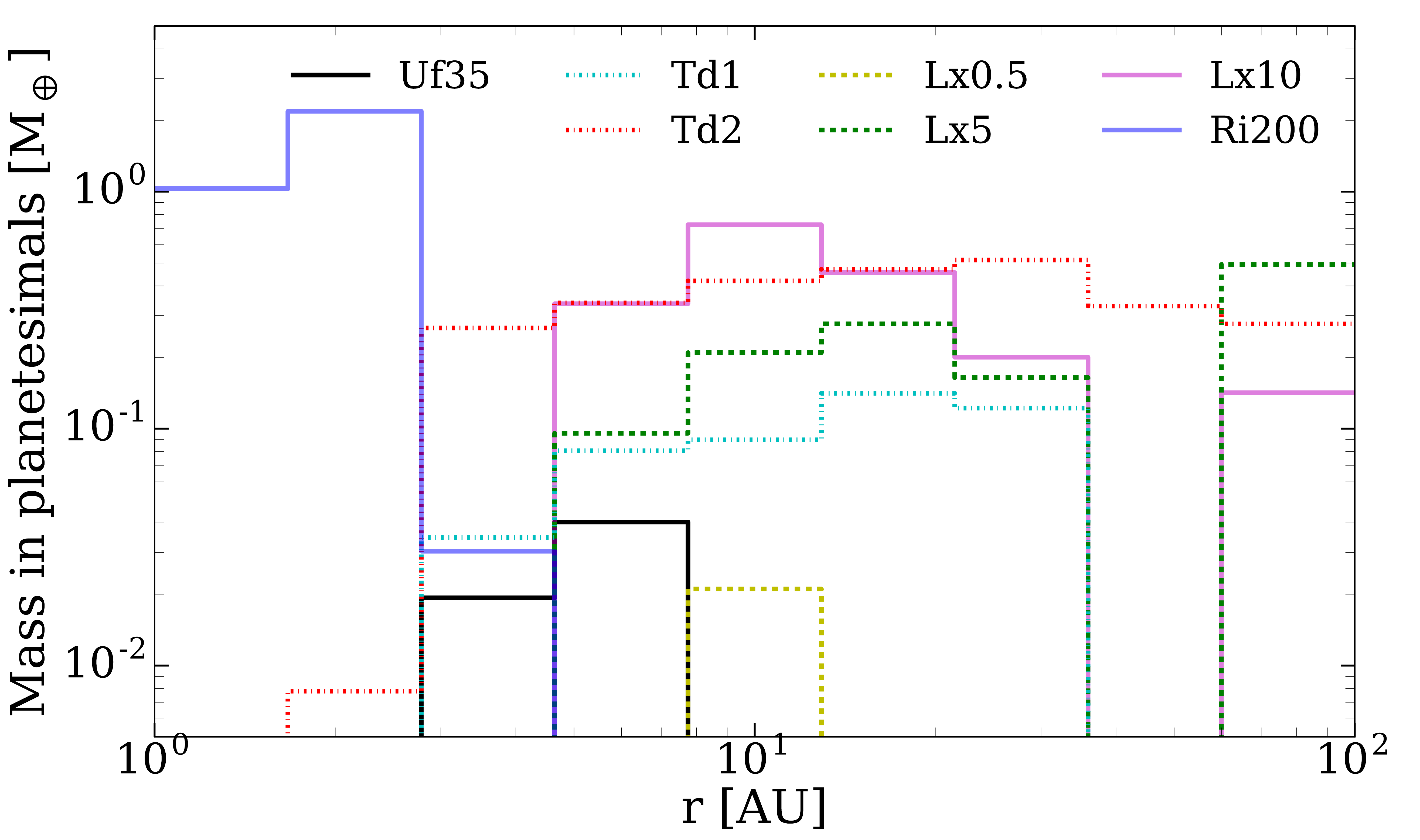}	
\label{fig:finalmass}
\caption{Total mass accumulated in planetesimals, {\mpl}, at the time of disc dispersal (taken as the time at which the photoevaporative hole reaches 100 AU) for models in Table~\ref{tab:models} in which the streaming instability is triggered. Note the binning here is different than in Table~\ref{tab:results}.}
\end{figure*}

Table~\ref{tab:results} summarises the mass accrued in planetesimals throughout the disc for the models summarised in Table~\ref{tab:models}. The streaming instability is never triggered in our fiducial model (i.e., no mass is accrued in planetesimals), and even models tuned to more favourable conditions for the triggering of the SI fall short of producing sufficient planetesimal masses to be relevant for the formation of terrestrial planets or giant planet cores. The latter is further problematic because giant planet cores that would form in an advancing pressure bump leading the photoevaporative hole would likely have insufficient time to acquire a gaseous envelope before the local gas supply is photoevaporated. This is in contrast to the results obtained by \citet{2017ApJ...839...16C}, which are summarised in the last row of Table~\ref{tab:results}. A discussion of the possible reasons for the discrepancy are in Section~\ref{sec:discussion}.

\begin{table*}
\noindent
\begin{tabular}{lccccccc}
\hline
Model           & \multicolumn{7}{c} {Mass converted into planetesimals, {\mpl} [{\me}]} \\ \cline{2-8}
    & Total  &  $<1$~AU & 1-3~AU & 3-10~AU & 10-30~AU & 30-100~AU & $>$100~AU \\
Fiducial      & 0     &   0      &    0      &        0    &      0   & 0& 0 \\
Uf5        & 0     &   0      &    0      &        0        &   0    & 0 & 0\\
Uf35   & 0.06  &   0      &    0      &        0.06 &    0     & 0 & 0\\
Td1       &  0.47 &   0      &   0    &       0.16  &  0.19 &   0.12& 0\\
Td2    & 2.62  &   0      &    0.04      &     0.79  &  0.96   & 0.84 & 0\\
Lx0.1 & 0        &   0      &    0      &   0      &    0      &  0& 0\\
Lx0.5 & 0.02   &   0      &    0      & 0.02  &  0 & 0 & 0\\
Lx5  & 1.24   &   0      &    0      &   0.30  &  0.28  &  0.66 & 0\\
Lx10 & 1.86    &   0      &    0      & 0.66  &  1.00  &  0.19& 0 \\
Ri200 & 3.24    &   0      &    3.23      & 0.01  &  0  &  0 & 0\\
\hline
CGJD17& 76.34 & 0.01 & 0.27 & 4.03 & 4.16 & 7.98 & 59.89 \\
\hline
\end{tabular}
\caption{Mass of solids converted into planetesimals via the streaming instability for each model in Table~\ref{tab:models}, binned in radius as in \citet{2017ApJ...839...16C}; that work's results are shown in the bottom row for comparison.}
\label{tab:results}
\end{table*}

Figure~\ref{fig:fiducial} summarises the results for our fiducial model, with snapshots of the  evolution of the gas and dust surface densities at major epochs, 10~kyr before and after gap opening at 3.22~Myr and when the inner edge of the outer disc has reached 100~AU at 4.1~Myr, shown in Figure~\ref{fig:fiducial}(a). The solid and dashed lines correspond to gas and dust surface density distributions, respectively. Once a photoevaporative gap opens in the gas disc, a pressure maximum is created that leads the outer edge of this gap (which quickly becomes a hole) and persists throughout the transition disc phase, sweeping outward as the disc is dispersed from the inside-out. Particles sensitive to radial drift are collected in this moving pressure \lq{}bump\rq{}, enhancing the local $Z$.

The maximum Stokes number reached in the disc (that of the largest grains) at the same epochs as in Figure~\ref{fig:fiducial}(a) are shown in Figure~\ref{fig:fiducial}(b). Our Stokes threshold to trigger the streaming instability, ${\rm St_{crit}} = 0.003$, is attained easily in much of the disc and always in the pressure bump that leads the photoevaporative gap/hole; it is instead insufficiently high $Z$ values that prevent triggering of the SI. This is a consequence of severe depletion of the dust disc by radial drift, which operates on a timescale much shorter than the gas disc lifetime (see Figure~\ref{fig:fiducial}(c)), such that
by the time of photoevaporative gap opening in the gas disc, the solids mass has fallen by two orders of magnitude. Even with sustained $Z$ enhancement in the pressure bump, this early depletion of the dust disc precludes attainment of $Z_{\rm crit}$ for the corresponding St and viscosity values. Figure~\ref{fig:fiducial}(d), a colour map for $Z$ over the final $\approx 25\%$ of the gas disc lifetime (during which photoevaporation dominates the disc evolution), demonstrates that while St values are sufficiently high ($0.01 \leq {\rm St} \leq 0.10$ within the overlaid black contour), $Z$ in the pressure bump fails to reach the threshold value.

By delaying the evolution of the dust disc we can artificially limit the loss of solids by radial drift, resulting in higher $Z$ in the late-stage pressure bump and subsequent triggering of the SI. Delaying the dust evolution by 1(2) Myr in our Td1(Td2) model produces 0.47(2.6) {\me} in planetesimals, distributed between 3(1) and 100 AU (see Table~\ref{tab:results}). Note however that delaying the dust insertion serves only as a proxy for some physical process that would strongly hinder radial drift on Myr timescales; we do not include any such process in this work in order to isolate the effects of photoevaporation. If dust traps were present and able to retain solids in the $1 - 100$ AU region of the disc, our delayed dust evolution models show that X-ray photoevaporation would then be efficient in triggering the instability.

\subsection{Sensitivity to dust fragmentation velocity}

Our fiducial model uses a fragmentation velocity $u_{\rm f} = 10$ {\ms}, though both lower and higher values are thought to be feasible as a consequence of the particles' surface energy or alternatively the size of the monomers \citep[see][]{2009ApJ...702.1490W,2015ApJ...798...34G,2013A&A...559A..62W,2016ApJ...818...16M}. Figure~\ref{fig:surface}(a) -- (b) show colour maps for the disc metallicity, $Z$, over the final $\sim$25\% of disc evolution for models with {\uf} of 5 and 35 {\ms}, respectively. Overlaid black and purple contours showing the regions in which Stokes numbers are $0.01 \leq {\rm St} \leq 0.10$ and $0.1 < {\rm St} \leq 1.0$, respectively. Comparing these results to Figure~\ref{fig:fiducial}(d) for our fiducial model, reducing the fragmentation velocity results in higher typical metallicity in the disc, as drift is reduced due to particles fragmenting before growing to larger sizes. As a consequence, however, the region with Stokes numbers higher than 0.01 is reduced. The criteria to trigger the streaming instability are thus not met with this lower {\uf}.

Conversely, for higher values of fragmentation velocity, $Z$ is on average lower throughout the disc, as particles are allowed to grow larger before fragmenting and are thus more susceptible to radial drift. The dimensionless stopping time is generally higher because of the larger fragmentation velocity, and Stokes numbers $0.1 < {\rm St} \leq 1.0$ are sustained in the pressure bump. This enables triggering of the streaming instability (red points in Figure~\ref{fig:surface}(b)) despite the lower average metallicity, though even with this effect strong as a result of the high fragmentation velocity $u_f = 35$ {\ms}, solids totaling only a few percent of an Earth mass are converted into planetesimals. While $Z$ and St profiles in the disc are thus altered by the chosen {\uf}, our results for planetesimal formation are largely insensitive to this choice over the range of reasonable fragmentation velocities. 

\subsection{Sensitivity to stellar X-ray luminosity}
The observed X-ray luminosities for low mass pre-main sequence stars show scatter over 2 orders of magnitude \citep{2005ApJS..160..390P}; since the wind mass loss rate scales directly with X-ray luminosity \citep{2012MNRAS.422.1880O}, the behaviour of a putative X-ray photoevaporation-induced streaming instability likely changes according to the properties of the central star. We explore this sensitivity by adjusting the X-ray luminosity of the host star by factors of 5 and 10 (see Table~\ref{tab:models}). For $L_{\rm X} =  10^{29}\ {\rm erg\ s^{-1}}$ (a one decade decrease from the fiducial value), the gas disc dispersal time exceeds 10 Myr, by which time only $\approx 0.5 {\rm M_\oplus}$ of dust remains in the entire disc, precluding any substantial planetesimal formation at the subsequent time of gap opening. When instead increasing the X-ray luminosity by one decade ($L_{\rm X} =  10^{31}\ {\rm erg\ s^{-1}}$), the higher mass loss rate opens a photoevaporative gap at $\approx 0.5$ Myr, at which time the solids mass in the disc is still $\approx 3\ {\rm M_\oplus}$. In this case $\approx 1.9\ {\rm M_\oplus}$ of planetesimals are formed via the streaming instability in the pressure bump as it leads the progressing gap/hole edge, with most of this formation occurring between $3 - 30$ AU (see Table~\ref{tab:results}). Note that the gas disc lifetimes we obtain with these low and high X-ray luminosities are on the high and low ends, respectively, of observations, such that these model results would likely apply to a minority of systems.

This motivates assessment of intermediate X-ray luminosity cases of $L_{\rm X} =  5 \times 10^{29}$ and $5 \times 10^{30}\ {\rm erg\ s^{-1}}$, shown in Figure~\ref{fig:surface}(c) and (d), respectively. As in Figure~\ref{fig:surface}(a) and (b), colour maps in (c) and (d) show $Z$ over the final $\approx 25\%$ of the gas disc lifetime, with overlaid black and purple contours bounding regions of $0.01 \leq {\rm St} \leq 0.10$ and $0.1 < {\rm St} \leq 1.0$, respectively. While the reduced stellar X-ray luminosity in model Lx0.5 (see Table~\ref{tab:models}) results in a longer dispersal timescale and thus a substantial depletion of the total dust mass by the time of gap opening, this model is able to trigger the SI in contrast to the higher {\lx} fiducial case. This is because the weaker {\lx} pushes the photoevaporative hole outward more slowly, thus moving the pressure bump outward more slowly, which allows a higher $Z$ enhancement at a given location to trigger the SI. The hole moves from 1 to 10 AU in the fiducial model in $\approx 300$ kyr, while the equivalent in the Lx0.5 model is $\approx 500$ kyr.
In contrast the dispersal timescale in model Lx5 is only $\approx 1.2$ Myr, though as in model Lx10, the higher dust mass remaining at the time of gap opening facilitates repeated crossing of the $Z_{\rm crit}$ threshold to trigger the SI. The final planetesimal mass in this case is $\approx 1.24\ {\rm M_\oplus}$, demonstrating modest sensitivity of planetesimal formation to the central star's {\lx}.

\subsection{Sensitivity to disc viscosity and outer radius}
\label{sec:ri200}
The drift timescale $\tau_{\rm drift} \propto r$, thus one avenue to retain solids is simply to consider a larger disc. In our simplified single $\alpha$ disc model, the initial disc size sets the viscous timescale for a given $\alpha$. Thus any adjustment of the initial disc size must be accompanied by a change of $\alpha$ to ensure that, together with photoevaporation, the resulting disc lifetime remains consistent with observations. In model Ri200 we increase the initial disc radius to 200 AU (cf. 18 AU in the fiducial model) and $\alpha$ to 0.01 ($7\times 10^{-4}$ in the fiducial model). Figure~\ref{fig:hialpha} shows results for this model in an analogous fashion to Figure~\ref{fig:fiducial} for the fiducial model. The longer drift timescale results in a depletion of the dust disc by only a factor of $\approx 5$ (cf. 2 dex in the fiducial model) by the time of gap opening, providing significantly more solids material for aggregation into planetesimals. Where in the fiducial model planetesimals are formed in a prominent pressure bump that leads the photoevaporative gap/hole, here instead the high $Z$ present near the peak of the X-ray-driven photoevaporation profile ($\approx 1 - 3$ AU) triggers the SI in this region before a gap is fully opened in the gas disc. As photoevaporation becomes relevant at $\approx 2.55$ Myr (see Figure~\ref{fig:hialpha}(d)), the depression it creates in the gas surface density (that becomes a full gap at $\approx 2.74$ Myr) is enough to enhance the local dust-to-gas ratio and trigger numerous SI events. Once a gap is fully opened, this effect persists briefly just beyond the outer edge of the gap (corresponding to the SI events shown as red dots in Figure~\ref{fig:hialpha}(d) along the hard boundary in white that represents the inner edge of the outer disc), and the larger $\alpha$ in this model causes the inner disc to drain particularly rapidly onto the star, resulting in subsequent direct irradiation of the outer disc and faster dispersal than in the fiducial case. This scenario yields a larger amount of planetesimals formed via the streaming instability than any of our other models, 3.24 {\me} formed predominantly between $\approx 1 - 3$ AU.

\section{Discussion and conclusions}
\label{sec:discussion}

The final mass formed in planetesimals as a function of radius in the disc for all models (Figure~\ref{fig:finalmass}) suggests that while X-rays can drive vigorous photoevaporative winds, their efficacy in producing the conditions to form planetesimals by the streaming instability is modest. These results support the conclusions of previous works that due to the late onset of dispersal by photoevaporative winds (compared to typical radial drift timescales), photoevaporation may only be relevant for the formation of debris discs, rather than a common mechanism for planet formation \citep{2014prpl.conf..475A}.

None of our models with standard initial conditions (initial disc radius $R_1 = 18$ AU and $\alpha = 7 \times 10^{-4}$) where dust is allowed to evolve from the beginning of disc evolution are able to produce sufficient mass in planetesimals to suggest this process is a major route to planet formation. While models with high X-ray luminosity do form planetesimal masses between a Mars and Earth mass in individual radial bins of Table~\ref{tab:results}, and our model with a higher fragmentation velocity {\uf} $= 35$ {\ms} stores Mercury's mass in planetesimals between 3 - 10 AU, this is still short of establishing photoevaporation as a major player in the formation of planetesimals by the streaming instability, particularly when considering the very low mid-plane viscosity imposed in our models and the failure to form sufficient planetesimal concentrations to explain the observed abundance of super-Earths/mini-Neptunes. 

Our conclusions are in contrast to the recent work of \citet{2017ApJ...839...16C}, who find that an FUV-driven wind \citep{2015ApJ...804...29G} is able to produce $\approx 76\ {\rm M_\oplus}$ of planetesimals before the gas disc is dispersed. While $\approx 16\ {\rm M_\oplus}$ of this forms between $ 1 - 100$ AU (Table~\ref{tab:results}), the rest is generated between $100 - 1000$ AU (primarily near the higher bound). While the global mass loss rates of X-ray driven photoevaporation are comparable to those of FUV driven photoevaporation, none of the X-ray photoevaporation models we explore forms a comparable amount of planetesimals, and we do not trigger the SI beyond 100 AU.

The primary difference between the two works' fiducial models is that we lose effectively the entire dust disc to radial drift onto the central star prior to the formation of the pressure bump by photoevaporation. The photoevaporative gap is opened at 3.22 Myr, at which time the total solids mass in the disc has fallen to $\approx 1\ {\rm M_\oplus}$ from an initial $\approx 232\ {\rm M_\oplus}$ (Figure~\ref{fig:fiducial}). The simulations of \citet{2017ApJ...839...16C}, by contrast, show a reduction in the dust disc mass by only a factor of a few at the onset of photoevaporation (see that work's Figure 3). Thus even in the first Myr of evolution, well before gap opening, the dust evolution is markedly different in the two works. One cause of the slower loss of solids in \citet{2017ApJ...839...16C} models is the assumed disc extent. However the discrepancy between our results persists even when we adopt a larger initial disc radius (200 AU) and $\alpha$ viscosity parameter (0.01; values comparable to those in \citet{2017ApJ...839...16C}) to slow the loss of solids by radial drift. In this case we produce $\approx 3\ {\rm M_\oplus}$ of planetesimals between $1 - 3$ AU, but we continue to form no planetesimals beyond 100 AU, where the bulk of production ($\approx 60\ {\rm M_\oplus}$ in \citet{2017ApJ...839...16C} lies.

That their models form a rich planetesimal population at large radii (beyond 100 and mostly near 1000 AU) and at early times (before 2 Myr) is due to retention of $\approx 100\ {\rm M_\oplus}$ of dust after the gas disc has dispersed, and \citet{2017ApJ...839...16C} state this is a direct consequence of their FUV-driven photoevaporation model; the rapid drop in gas surface density at large radii due to photoevaporation stalls radial transport in the outermost disc. We thus conclude that, assuming no inconsistencies in the implementation of the dust evolution model between the two codes, the difference in results for the efficacy of photoevaporation to form planetesimals by the streaming instability is due principally to the choice of radial photoevaporation profile used in the disc (rather than differences in the total photoevaporative mass loss rates or the initial configuration of the viscously evolving disc). Unlike the X-ray driven wind profile adopted in this work, the wind profile in the FUV-driven case presented by \citet{2015ApJ...804...29G} and used in \citet{2017ApJ...839...16C} is efficient at removing gas at disc radii beyond 100 AU. This results in a gas depleted dust disc extending to $\approx 1000$ AU that persists as the gas disc is dispersed over a few Myr. This prediction seems at odds with observations that show larger gas than dust discs \citep[e.g.,][]{HD163296}.

While it is beyond the scope of this paper to compare the prevailing photoevaporation models and their individual successes in reproducing observations (for recent work on this see the reviews \citet{2014prpl.conf..475A} and \citet{2017RSOS....470114E}), a comparison of our results with those in \citet{2017ApJ...839...16C} does make clear that the relevance of photoevaporation to the formation of planetesimals via mechanisms such as the streaming instability depends heavily on the choice of radial photoevaporation profile in the disc. While we find that a photoevaporation-induced streaming instability plays a marginal role in the formation of planetary systems, the role of photoevaporation in planetesimal formation remains uncertain until photoevaporation models can be more tightly constrained with observational diagnostics.

\section*{Acknowledgements}
We thank the anonymous referees for suggestions that helped improving the clarity of our paper. We thank Andrew Youdin and Cathie Clarke for the useful suggestions and discussions. J.J. thanks T. Nealer for her comments on the work. This work was funded by the Deutsche Forschungsgemeinschaft (DFG, German Research Foundation) - Ref no. FOR 2634/1. GR acknowledges support from the DISCSIM project, grant
agreement 341137 funded by the European Research Council under
ERC-2013-ADG. T.B. acknowledges funding from the European Research Council (ERC) under the European Union's Horizon 2020 research and innovation programme under grant agreement No 714769. This research was supported by the Munich Institute for Astro- and Particle Physics (MIAPP) of the DFG Cluster of Excellence \lq{}Origin and Structure of the Universe.\rq{}
% * <ercolano@gmx.de> 2017-08-31T10:13:09.561Z:
%
% ^.

%%%%%%%%%%%%%%%%%%%%%%%%%%%%%%%%%%%%%%%%%%%%%%%%%%


\bibliographystyle{mnras}
\bibliography{references}

%%%%%%%%%%%%%%%%%%%%%%%%%%%%%%%%%%%%%%%%%%%%%%%%%%

%%%%%%%%%%%%%%%%% APPENDICES %%%%%%%%%%%%%%%%%%%%%

%\appendix

%\section{}

%%%%%%%%%%%%%%%%%%%%%%%%%%%%%%%%%%%%%%%%%%%%%%%%%%

% Don't change these lines
\bsp	% typesetting comment
\label{lastpage}
	\end{document}